\documentclass[12pt,preprint,preprintnumbers,nofootinbib,groupedaddress,superscriptaddress,amsmath,amssymb]{revtex4}

\usepackage{graphicx}
\usepackage{dcolumn}
\usepackage{bm}
\usepackage{amssymb}
\usepackage{amsmath}
\usepackage{epsfig}    
\usepackage{color}
\usepackage{hhline}

\def\be{\begin{equation}}
\def\ee{\end{equation}}

\def\nn{\nonumber}

\numberwithin{equation}{section}

\begin{document}
{\begin{flushright}{ APCTP Pre2019 - 022}\end{flushright}}

\title{Neutrino mass model with a modular $S_4$ symmetry}

\author{Hiroshi Okada}
\email{hiroshi.okada@apctp.org}
\affiliation{Asia Pacific Center for Theoretical Physics (APCTP) - Headquarters San 31, Hyoja-dong,
Nam-gu, Pohang 790-784, Korea}

\author{Yuta Orikasa}
\email{Yuta.Orikasa@utef.cvut.cz}
\affiliation{Institute of Experimental and Applied Physics, 
Czech Technical University in Prague, 
Husova 240/5, 110 00 Prague 1, Czech Republic}

\date{\today}

\begin{abstract}
We propose a predictive lepton model under a modular $S_4$ symmetry,
where the neutrino mass matrix arises from a radiative seesaw at one-loop level.
The tree-level mass matrix is forbidden by well-assigned modular weights, which also play an important role in stabilizing dark matter candidate due to a remnant $Z_2$ symmetry even after breaking the modular symmetry.
Supposing three families of the Majorana neutrinos, right-handed charged-leptons and left-handed charged-leptons to be embedded respectively into singlet, doublet, and triplet under $S_4$, we obtain the predictive mass matrices in the normal hierarchy.
Then, we show our numerical results such as phases, mixings, and neutrino masses, applying $\chi^2$ analysis.
We also demonstrate two sample points, imposing on minimizing $\chi^2$ and best fit value of $\delta_{CP}^\ell$ of $195^\circ$.
\end{abstract}
\maketitle
\newpage

\section{Introduction}
 Neutrino and dark matter (DM) physics are big issues to be solved beyond the standard model (SM),
 even though SM successfully describes a lot of phenomenologies in high energy physics.
Radiative seesaw models are one of the attractive scenarios not only to explain both but also make a correlation 
between them. The first approach is achieved by Ref.~\cite{Ma:2006km}, in which neutrino mass matrix is given at one-loop level, and a Majorana fermion DM or an inert scalar DM is included in the neutrino mass loop.
It is also an important issue to resolve the flavor puzzles such as lepton flavor violations (LFVs), Z boson decays, flavor changing neutral currents enlightened by a lot of experimental results.
These issues often arise from rather large Yukawa couplings that frequently appear on the radiative seesaw models,
even though it also depends on structures of Yukawa matrices. 
If there exists a flavor symmetry such that their matrices are uniquely determined, it might provide powerful hints to the flavor physics.

Recently, modular flavor symmetries have been proposed~\cite{Feruglio:2017spp, deAdelhartToorop:2011re}
to provide more predictions to the quark and lepton sector due to Yukawa couplings with a representation of a group.  
Their typical groups are found in basis of  the $A_4$ modular group \cite{Feruglio:2017spp, Criado:2018thu, Kobayashi:2018scp, Okada:2018yrn, Nomura:2019jxj, Okada:2019uoy, deAnda:2018ecu, Novichkov:2018yse, Nomura:2019yft, Okada:2019mjf, Nomura:2019lnr}, $S_3$ \cite{Kobayashi:2018vbk, Kobayashi:2018wkl, Kobayashi:2019rzp, Okada:2019xqk}, $S_4$ \cite{Penedo:2018nmg, Novichkov:2018ovf, Kobayashi:2019mna}, $A_5$ \cite{Novichkov:2018nkm, Ding:2019xna}, larger groups~\cite{Baur:2019kwi, Ding:2020msi}, multiple modular symmetries~\cite{deMedeirosVarzielas:2019cyj}, and double covering of $A_4$~\cite{Liu:2019khw} in which  masses, mixings, and CP phases for quark and lepton are predicted.~\footnote{Several reviews are helpful to understand the whole idea~\cite{Altarelli:2010gt, Ishimori:2010au, Ishimori:2012zz, Hernandez:2012ra, King:2013eh, King:2014nza, King:2017guk, Petcov:2017ggy}.}
Also, a systematic approach to understand the origin of CP transformations has been recently achieved by ref.~\cite{Baur:2019iai}.

In this paper, we introduce a modular $S_4$ symmetry in the lepton sector, and the neutrino mass matrix is arisen via radiative seesaw at one-loop level.
Supposing three families of the Majorana neutrinos, right-handed charged-leptons and left-handed charged-leptons to be embedded respectively into singlet, doublet, and triplet under $S_4$, we obtain the predictive mass matrices.
Then, we show our numerical results such as phases, mixings, and neutrino masses, applying $\chi^2$ analysis.
Instead of an additional symmetry such as $Z_2$ to stabilize DM, the model has a remnant symmetry of modular symmetry. This is why we have an appropriate 
DM candidate.

This paper is organized as follows.
In Sec.~\ref{sec:realization},   we explain our model and formulate mass matrices, LFVs, and so on
under the modular $S_4$ symmetry. 
Then, we show numerical analyses for normal hierarchy (NH) and inverted hierarchy (IH) and discuss our predictions.
We summarize and conclude in Sec.~\ref{sec:conclusion}. In appendix, we note the correspondence between confidential level(CL) and $\chi^2$, depending on the number of degrees of freedom for observables.

\begin{center} 
\begin{table}[tb]
\begin{tabular}{|c||c|c|c|c|c|c|c||c|c|}\hline
&\multicolumn{5}{c||}{ Fermions} & \multicolumn{2}{c||}{Bosons} \\\hline
  & ~$\bar L_{L_{e,\mu,\tau}}$~  & ~$e_{R_e}$~& ~$\ell_{R}\equiv(e_{R_{\mu}},e_{R_{\tau}})^T$~& ~$N_{R}\equiv(N_{R_{1}},N_{R_{2}})^T$~ & ~$N_{R_3}$~ & ~$H$~  & ~$\eta^*$~
  \\\hline 
 $SU(2)_L$ & $\bm{2}$    & $\bm{1}$ & $\bm{1}$    & $\bm{1}$  & $\bm{1}$ & $\bm{2}$ & $\bm{2}$    \\\hline 
$U(1)_Y$ & $\frac12$    & $-1$& $-1$   & $0$  & $0$  & $\frac12$& $-\frac12$      \\\hline
 $S_4$ & $3$ & $1$ & $2$ & $2$ & $1$ & $1$ & $1$     \\\hline
 $-k$ &   $-2$ & $-2$ & $-2$ & $-1$ & $-1$ & $0$ & $-3$   \\\hline
\end{tabular}
\caption{Field contents of fermions and bosons
and their charge assignments under $SU(2)_L\times U(1)_Y\times S_4$ in the lepton and boson sector, 
where $-k$ is the number of modular weight
and the quark sector is the same as the SM.}
\label{tab:fields}
\end{table}
\end{center}


\begin{center} 
\begin{table}[tb]
\begin{tabular}{|c||c|c|c|c|c|c|c|c|c||}\hline
 &\multicolumn{7}{c|}{Couplings}  \\\hline
  & ~$Y^{(2)}_{\bf2}$~& ~$Y^{(4)}_{\bf3}$~  & ~$Y^{(4)}_{\bf3'}$ & ~$Y^{(6)}_{\bf 1}$& ~$Y^{(6)}_{\bf3}$~  & ~$Y^{(6)}_{\bf3'_1}$ & ~$Y^{(6)}_{\bf3'_2}$ \\\hline
 $S_4$ & ${\bf2}$ & ${\bf3}$ & ${\bf 3'}$ &${\bf 1}$ & ${\bf3}$ & ${\bf3'}$ & ${\bf3'}$  \\\hline
 $-k$ & $2$ & $4$& $4$ & $6$  & $6$ & $6$  & $6$   \\\hline
\end{tabular}
\caption{Modular $S_4$ representations for Yukawa couplings.}
\label{tab:couplings}
\end{table}
\end{center}

\section{ Model} 
\label{sec:realization}
Here, we describe our scenario based on Ma model, where 
field contents are exactly the same as Ma model~\cite{Ma:2006km}. 
We introduce the Majorana fermions to be embedded into singlet and doublet under $S_4$ and 
scalar bosons to be embedded into doublet under $SU(2)_L$. 
The singlet right-handed fermion is required because the neutrino oscillation data can't be explained in the model without singlet fermion. 
The {$S_4$ representation} and modular weight
are given by Tab.~\ref{tab:fields}, while the ones of Yukawa couplings are given by Tab.~\ref{tab:couplings}.
The model has a remnant $Z_2$ symmetry. 
Particles with odd modular weight have odd parity and particles with even modular weight have even parity. 
Under these symmetries, one writes renormalizable Lagrangian as follows:
\begin{align}
-&{\cal L}_{Lepton} 
\supset
\alpha_\ell (Y^{(4)}_{\bf 3}\otimes\bar L_L\otimes e_R)_{\bf1}H
+\beta_\ell (Y^{(4)}_{\bf 3}\otimes\bar L_L\otimes \ell_R)_{\bf1}H
+\gamma_\ell (Y^{(4)}_{\bf 3'}\otimes\bar L_L\otimes \ell_R)_{\bf1}H
\nn\\
&
+\alpha_\eta (Y^{(6)}_{\bf 3}\otimes\bar L_{L}\otimes N_{R})_{\bf1}\tilde\eta
+ \left( (\beta_{\eta_1}Y^{(6)}_{\bf 3'_1} + \beta_{\eta_2} Y^{(6)}_{\bf 3'_2})\otimes\bar L_{L}\otimes N_{R}\right)_{\bf1}\tilde\eta
+\gamma_\eta (Y^{(6)}_{\bf 3} \otimes\bar L_{L}\otimes N_{R_3})_{\bf1}\tilde\eta
\nn\\
&+ M_0 (Y^{(2)}_{\bf 2} \otimes\bar N^C_{R}\otimes N_{R})_{\bf1}
+ M_1 (Y^{(2)}_{\bf 2} \otimes\bar N^C_{R}\otimes N_{R_3})_{\bf1}+ {\rm h.c.}, \label{eq:lag-lep}
\end{align}
where $\tilde\eta\equiv i\sigma_2 \eta^*$, $\sigma_2$ being the second Pauli matrix. 

The  modular forms with the lowest weight 2 are given by $Y^{(2)}_{\bf2}\equiv  [y_1,y_2]^T$ 
and they respectively transform
as a doublet and a triplet under $S_4$ that are written in terms of Dedekind eta-function  $\eta(\tau)$ and its derivative \cite{Novichkov:2019sqv}:
\begin{eqnarray} 
\label{eq:Y-S3}
y_1(\tau) &=& \frac{i}{8} \left( 8 \frac{\eta'(\tau+\frac12)}{\eta(\tau+\frac12)}  + 32 \frac{\eta'(4\tau)}{\eta(4\tau)}  
- \frac{\eta'(\frac\tau4)}{\eta(\frac\tau4)}- \frac{\eta'(\frac{\tau+1}4)}{\eta(\frac{\tau+1}4)}
- \frac{\eta'(\frac{\tau+2}4)}{\eta(\frac{\tau+2}4)} - \frac{\eta'(\frac{\tau+3}4)}{\eta(\frac{\tau+3}4)}  \right), \nn \\
y_2(\tau) &=& \frac{ i \sqrt3}{8} \left(\frac{\eta'(\frac\tau4)}{\eta(\frac\tau4)} - \frac{\eta'(\frac{\tau+1}4)}{\eta(\frac{\tau+1}4)}
 + \frac{\eta'(\frac{\tau+2}4)}{\eta(\frac{\tau+2}4)} - \frac{\eta'(\frac{\tau+3}4)}{\eta(\frac{\tau+3}4)}  \right)
 \nn \\
y_3(\tau) &=& i \left(\frac{\eta'(\tau+\frac12)}{\eta(\tau+\frac12)}  - 4 \frac{\eta'(4\tau)}{\eta(4\tau)}  \right),\nn\\
y_4(\tau) &=& \frac{i}{4\sqrt2} \left( - \frac{\eta'(\frac\tau4)}{\eta(\frac\tau4)}+i \frac{\eta'(\frac{\tau+1}4)}{\eta(\frac{\tau+1}4)}
+\frac{\eta'(\frac{\tau+2}4)}{\eta(\frac{\tau+2}4)} - i \frac{\eta'(\frac{\tau+3}4)}{\eta(\frac{\tau+3}4)}  \right), \nn \\
y_5(\tau) &=&  \frac{i}{4\sqrt2} \left( - \frac{\eta'(\frac\tau4)}{\eta(\frac\tau4)} - i \frac{\eta'(\frac{\tau+1}4)}{\eta(\frac{\tau+1}4)}
 + \frac{\eta'(\frac{\tau+2}4)}{\eta(\frac{\tau+2}4)} +i \frac{\eta'(\frac{\tau+3}4)}{\eta(\frac{\tau+3}4)}  \right)
\label{Yi}.
\end{eqnarray}
%
Then, higher weights are constructed by multiplication rules of $S_4$,
and one finds the following couplings:
\begin{align}
&Y^{(4)}_{\bf3}=
\left[\begin{array}{c}
-2y_2y_3 \\ 
\sqrt3 y_1 y_5+ y_2 y_4  \\ 
\sqrt3 y_1 y_4+ y_2 y_5  \\ 
\end{array}\right],\quad
Y^{(4)}_{\bf3'}=
\left[\begin{array}{c}
2y_1y_3 \\ 
\sqrt3 y_2 y_5 - y_1 y_4  \\ 
\sqrt3 y_2 y_4- y_1 y_5  \\ 
\end{array}\right],\\
&
Y^{(6)}_{\bf3}=
\left[\begin{array}{c}
y_1 (y_4^2-y_5^2) \\ 
y_3 (y_1 y_5+\sqrt3 y_2 y_4) \\ 
-y_3 (y_1 y_4+\sqrt3 y_2 y_5) \\ 
\end{array}\right],\quad
Y^{(6)}_{\bf3'_1}=
(y_1^{2}+y_2^{2})
\left[\begin{array}{c}
y_3 \\ 
y_4 \\ 
y_5 \\ 
\end{array}\right],\quad
Y^{(6)}_{\bf3'_2}=
\left[\begin{array}{c}
y_2 (y_5^2-y_4^2) \\ 
y_3 (y_2 y_5-\sqrt3 y_1 y_4) \\ 
y_3 (y_2 y_4-\sqrt3 y_1 y_5) \\ 
\end{array}\right].
\end{align}

After the electroweak spontaneous symmetry breaking,  the charged-lepton mass matrix is given by
\begin{align}
m_\ell&= \frac {v_H}{\sqrt{2}}
\left[\begin{array}{ccc}
\alpha_\ell Y_1 & \beta_\ell Y_1 & -\gamma_\ell Y'_1 \\ 
\alpha_\ell Y_3 & -\frac{1}2 \beta_\ell Y_3 + \frac{\sqrt3}{2}\gamma_\ell Y'_2 & \frac{\sqrt3}2\beta_\ell Y_2 + \frac{1}{2}\gamma_\ell Y'_3  \\ 
\alpha_\ell Y_2 & -\frac{1}2 \beta_\ell Y_2 + \frac{\sqrt3}{2}\gamma_\ell Y'_3 & \frac{\sqrt3}2\beta_\ell Y_3 + \frac{1}{2}\gamma_\ell Y'_2  \\ \end{array}\right], 
\end{align}
where {$\langle H\rangle\equiv [0, v_H/\sqrt2]^T$}, $Y^{(4)}_{\rm 3}\equiv [Y_1,Y_2,Y_3]^T$, and $Y^{(4)}_{\rm 3'}\equiv [Y'_1,Y'_2,Y'_3]^T$.
Then the charged-lepton mass eigenstate can be found by $|D_\ell|^2\equiv V_{e_L} m_\ell m^\dag_\ell V_{e_L}^\dag$.
In our numerical analysis below, we fix the free parameters $\alpha_\ell,\beta_\ell,\gamma_\ell$ to fit the three charged-lepton masses after giving all the numerical values, by applying the relations:
\begin{align}
&{\rm Tr}[m_\ell {m_\ell}^\dag] = |m_e|^2 + |m_\mu|^2 + |m_\tau|^2,\\
&{\rm Det}[m_\ell {m_\ell}^\dag] = |m_e|^2  |m_\mu|^2  |m_\tau|^2,\\
&({\rm Tr}[m_\ell {m_\ell}^\dag])^2 -{\rm Tr}[(m_\ell {m_\ell}^\dag)^2] =2( |m_e|^2  |m_\nu|^2 + |m_\mu|^2  |m_\tau|^2+ |m_e|^2  |m_\tau|^2 ).
\end{align}

 The right-handed neutrino mass matrix is given by
\begin{align}
{\cal M_N} &=
M_0
\left[\begin{array}{ccc}
 -y_1 & y_2  & 0\\ 
y_2 & y_1 & 0   \\ 
0 & 0 & 0   \\ 
\end{array}\right]
+
M_1
\left[\begin{array}{ccc}
 0 & 0  & y_1\\ 
0 & 0 & y_2   \\ 
y_1 & y_2 & 0   \\ 
\end{array}\right]. 
\label{eq:mn}
\end{align}
The heavy Majorana mass matrix is diagonalized by  a unitary matrix $V$ as follows: $D_N\equiv V^* {\cal M_N}V^\dag$,
where $N_R\equiv V^\dag \psi_R$, $\psi_R$ being mass eigenstate.

The Dirac Yukawa matrix is given by
\begin{align}
y_\eta &=
\left[\begin{array}{ccc}
\alpha_\eta Y^6_{1} 
&-\beta_{\eta_1} Y^{\prime 6}_{1} - \beta_{\eta_2} Y^{\prime\prime 6}_{1}  
& \gamma_\eta Y^6_1 
\\ 
-\frac{1}{2}\alpha_\eta Y^6_{3} +\frac{\sqrt3}2 ( \beta_{\eta_1} Y^{\prime\prime 6}_{2} + \beta_{\eta_2} Y^{\prime\prime 6}_{2})
&\frac{\sqrt3}{2}\alpha_\eta Y_{2} +\frac{\sqrt1}2 ( \beta_{\eta_1} Y^{\prime\prime 6}_{3} + \beta_{\eta_2} Y^{\prime\prime 6}_{3})   
& \gamma_\eta Y^6_3   
\\ 
-\frac{1}{2}\alpha_\eta Y_{2} +\frac{\sqrt3}{2}( \beta_{\eta_1} Y^{\prime\prime 6}_{3} + \beta_{\eta_2} Y^{\prime\prime 6}_{3})  
&\frac{\sqrt3}{2}\alpha_\eta Y^6_{3} +\frac{\sqrt1}{2}( \beta_{\eta_1} Y^{\prime\prime 6}_{2} + \beta_{\eta_2} Y^{\prime\prime 6}_{2})    
& \gamma_\eta Y^6_2   
\\ 
\end{array}\right], 
\label{eq:mn}
\end{align}
where $Y^{(6)}_{\rm 3}\equiv [Y^6_1,Y^6_2,Y^6_3]^T$ ,
$Y^{(6)}_{\rm 3'_1}\equiv [Y^{\prime 6}_1,Y^{\prime 6}_2,Y^{\prime 6}_3]^T$ and 
$Y^{(6)}_{\rm 3'_2}\equiv [Y^{\prime\prime 6}_1,Y^{\prime\prime 6}_2,Y^{\prime\prime 6}_3]^T$.

{\it Scalar potential} is given by
\begin{align}
{\cal V} = 
& -\mu_H^2 |H|^2 +\mu^2_\eta |\eta|^2
 + \lambda_H|H|^4+ \lambda_\eta  |\eta|^4 
 +\lambda_{H\eta} |H|^2|\eta|^2
\nn\\
& +\lambda_{H\eta}' |H^\dag\eta|^2
 +\frac12 \lambda_{H\eta}'' [(H^\dag\eta)^2+ {\rm h.c.}], 
 \label{eq:pot}
\end{align}
where $\lambda_{H\eta}''$ includes $Y_{\bf 1}^{(6)}$ factor and $\mu_\eta^2, \lambda_\eta, \lambda_{H\eta}, \lambda_{H\eta}'$ 
include $1/(-i \tau + i\bar{\tau})^n$ factor.
Scalar masses are given by 
\begin{eqnarray}
 m_h^2 &=& 2 \lambda_H v^2 ,
 \\
 m_R^2 &=& \mu_\eta^2+\frac12 \left( \lambda_{H\eta} + \lambda_{H\eta}' + \lambda_{H\eta}'' \right) v_H^2, 
\\
 m_I^2 &=& \mu_\eta^2+\frac12 \left( \lambda_{H\eta} + \lambda_{H\eta}' - \lambda_{H\eta}'' \right) v_H^2, 
 \\
 m_{\eta^\pm}^2 &=& \mu_\eta^2+\frac12 \lambda_{H\eta} v_H^2,
\end{eqnarray} 
where $m_{R(I)}$ is a mass of the real (imaginary) component of $\eta^0$.

{\it Lepton flavor violations} also arise from $y_\eta$ as~\cite{Baek:2016kud, Lindner:2016bgg}
\begin{align}
&{\rm BR}(\ell_i\to\ell_j\gamma)\approx\frac{48\pi^3\alpha_{em}C_{ij}}{G_F^2 (4\pi)^4}
\left|\sum_{\alpha=1}^3 Y_{\eta_{j\alpha}} Y^\dag_{\eta_{\alpha i}} F(D_{N_\alpha},m_{\eta^\pm})\right|^2,\\
&F(m_a,m_b)\approx\frac{2 m^6_a+3m^4_am^2_b-6m^2_am^4_b+m^6_b+12m^4_am^2_b\ln\left(\frac{m_b}{m_a}\right)}{12(m^2_a-m^2_b)^4},
\end{align}
where $Y_\eta\equiv y_\eta V^\dag$ $C_{21}=1$, $C_{31}=0.1784$, $C_{32}=0.1736$, $\alpha_{em}(m_Z)=1/128.9$, and $G_F=1.166\times10^{-5}$ GeV$^{-2}$.
The experimental upper bounds are given by~\cite{TheMEG:2016wtm, Aubert:2009ag,Renga:2018fpd}
\begin{align}
{\rm BR}(\mu\to e\gamma)\lesssim 4.2\times10^{-13},\quad 
{\rm BR}(\tau\to e\gamma)\lesssim 3.3\times10^{-8},\quad
{\rm BR}(\tau\to\mu\gamma)\lesssim 4.4\times10^{-8},\label{eq:lfvs-cond}
\end{align}
which will be imposed in our numerical calculation.

{\it Neutrino mass matrix}  is given by a combination of canonical seesaw at tree-level and radiative seesaw at one-loop level by
\begin{align}
 m_{\nu_{ij}} &=   \sum_{\alpha=1}^3 \left[
\frac{Y_{\eta_{i\alpha}} M_{N_{\alpha}} Y^T_{\eta_{\alpha j}}}{2(4\pi)^2}
\left(\frac{m_R^2}{m_R^2- M^2_{N_{\alpha}}}\ln\left[\frac{m_R^2}{M^2_{N_{\alpha}}}\right]
- \frac{m_I^2}{m_I^2-{M^2_N}_{\alpha}}\ln\left[\frac{m_I^2}{M^2_{N_{\alpha}}}\right]
\right)
\right] ~\nn\\
&\simeq \frac{\lambda_{H\eta}''   v_H^2}{ 2 (4\pi)^2}\sum_{\alpha =1}^{3} 
\frac{Y_{\eta_{i\alpha}} M_{N_{\alpha}} Y^T_{\eta_{\alpha j}} }{m_0^2-M^2_{N_\alpha}}
\left[ 1- 
\frac{ M_{N_{\alpha}}^2 }{m^2_0-M^2_{N_\alpha}} \ln \frac{m^2_0}{M^2_{N_\alpha}}  \right] ,
\label{eq:numass}
\end{align}
where we assume to be $ \lambda_{H\eta}''  v_H^2 = m^2_R-m^2_I << m_0^2\equiv (m_R^2+m_I^2)/2$ in the above second line. 
The neutrino mass matrix is diagonalized by a unitary matrix $U_{\nu}$ as $U_{\nu}m_\nu U^T_{\nu}=$diag($m_{\nu_1},m_{\nu_2},m_{\nu_3}$)$\equiv D_\nu$, 
where Tr$[D_{\nu}] \lesssim$ 0.12 eV is given by the recent cosmological data~\cite{Aghanim:2018eyx, Vagnozzi:2017ovm}.
The two mass squared differences are also measure by experiments and they are defined by 
\begin{align}
& \rm NH: \Delta m^2_{\rm sol}= m^2_{\nu_2}-m^2_{\nu_1},\ \Delta m^2_{\rm atm}= m^2_{\nu_3}-m^2_{\nu_1} ,\\
& \rm IH: \Delta m^2_{\rm sol}= m^2_{\nu_2}-m^2_{\nu_1},\ \Delta m^2_{\rm atm}= m^2_{\nu_2}-m^2_{\nu_3}.
\end{align}
We use $\lambda_{H\eta}''$ given by the following relation: 
\begin{eqnarray}
 \lambda_{H\eta}'' = 
 \sqrt{
 \frac{\Delta m_{\rm sol}^2}{ \tilde{m}^2_{\nu_2} - \tilde{m}^2_{\nu_1}}
 } ,
\end{eqnarray}
where $\tilde{m}_{\nu_i} \equiv m_{\nu_i} / \lambda_{H\eta}''$.
Then, one finds $U_{PMNS}=V^\dag_{eL} U_\nu$.
Mixing angles are given in terms of the component of $U_{PMNS}$ as follows:
\begin{align}
\sin^2\theta_{13}=|(U_{PMNS})_{13}|^2,\quad 
\sin^2\theta_{23}=\frac{|(U_{PMNS})_{23}|^2}{1-|(U_{PMNS})_{13}|^2},\quad 
\sin^2\theta_{12}=\frac{|(U_{PMNS})_{12}|^2}{1-|(U_{PMNS})_{13}|^2}.
\end{align}

In the model, NH is favored by the structures of Yukawa couplings and Majorana mass matrix. 
There are two light neutrino masses and one heavy mass.(see appendix A) 
Therefore the structures naturally realize NH of neutrino masses. 

Also, the effective mass for the neutrinoless double beta decay is given by
\begin{align}
\langle m_{ee}\rangle=|m_{\nu_1} \cos^2\theta_{12} \cos^2\theta_{13}+m_{\nu_2} \sin^2\theta_{12} \cos^2\theta_{13}e^{i\alpha_{21}}
+m_{\nu_3} \sin^2\theta_{13}e^{i(\alpha_{31}-2\delta_{CP})}|,
\end{align}
where its observed value could be measured by KamLAND-Zen in future~\cite{KamLAND-Zen:2016pfg}.

\subsection{Numerical analysis}
Here, we demonstrate  numerical analysis to find predictions as well as reproduce the current experimental results,
where we suppose the DM candidate is an imaginary component of inert scalar $\eta$; $\eta_I$, in which we simply assume {$m_{\eta^\pm} \approx m_I$} to evade the oblique parameters.
In this case, the mass of DM is within $534\pm8.5$ GeV~\cite{Hambye:2009pw} to satisfy the relic density that arises from the kinetic term only. Here, we work on this range.

Our input parameters are three mass parameters; $m_R$, $M_0$, $M_1$ and six dimensionless parameters $\tau$, $\alpha_D, \lambda_{H\eta}''$, $\alpha_D,\beta_D,\gamma_D$, where we work on the following ranges for both cases:
\begin{align}
& \tau = [-1.5, 1.5] +i [0.5, 2],\ [|\alpha_D|, \lambda_{H\eta}'']=[1,10]\times10^{-7},
\ [|\alpha_\eta|,|\beta_\eta|,|\gamma_\eta|]=[0.001,1],
\nn\\
& m_{R}=[525.5, 542.5]\ {\rm GeV},\
M_0=[10^3, 10^6] \ {\rm GeV}, \ 
M_0 \le M_1\le 10\times M_0.\nn
\end{align}
Notice here $m_I=\sqrt{m_R^2+2  \lambda_{H\eta}'' v_H^2}$, and $ \lambda_{H\eta}''$ is assumed to be a real parameter.
Next, we will show some plots in terms of the classification of $\chi$ square analysis within the range of 1-2$\sigma$ which is represented by green color, 2-3$\sigma$ which is represented by yellow color, and 3-5$\sigma$ which is represented by red color, referring to NuFIT 5.0~\cite{Esteban:2020cvm}. In the present work, we adopt the accuracy of $\chi^2$ for five well known dimensionless observables such as $\Delta m_{\rm atm}^2$, $\sin^2 \theta_{23}$, $\sin^2 \theta_{13}$, $\Delta m_{\rm sol}^2$, and $\sin^2 \theta_{12}$. Notice here that the masses of charged-lepton can precisely be fitted by $\alpha_\ell, \beta_\ell, \gamma_\ell$.
Thus, we do not include these masses in the $\chi^2$ analysis.
\begin{figure}[tb]\begin{center}
\includegraphics[width=80mm]{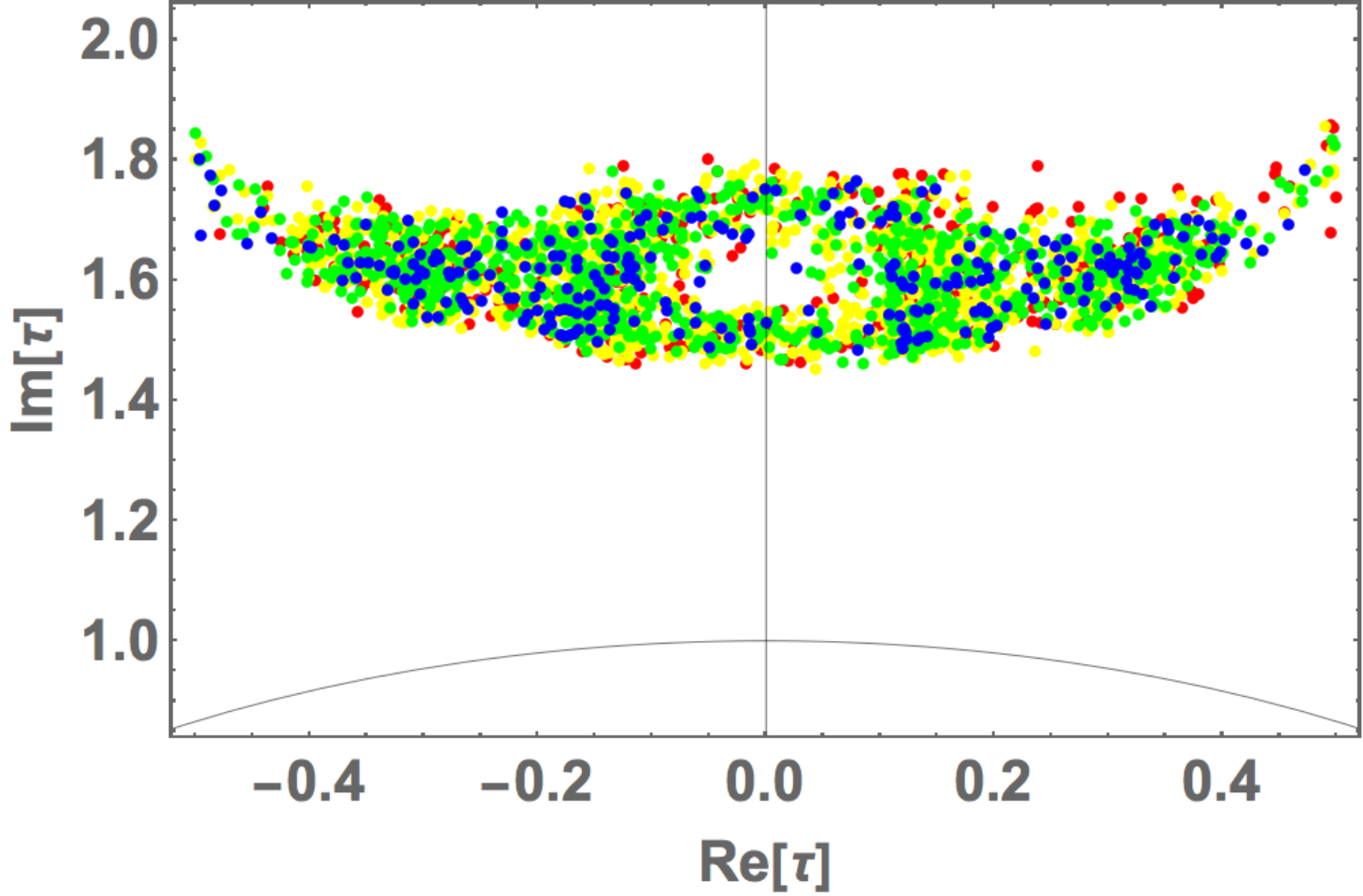}
\caption{Allowed region of $\tau$ in case of NH.
In the $\chi^2$ analysis, the blue color represents $\le1\sigma$, green 1-2$\sigma$, yellow 2-3$\sigma$, and red 3-5$\sigma$. The black solid line is the boundary of the fundamental domain at $|\tau|=1$.}   
\label{fig:tau}\end{center}\end{figure}

In case of NH, the lightest $D_N$ is heavier than 500 TeV. 
These suggest that they are totally safe for LFV constraints.
In case of IH, we need fine-tuned parameters and the LFV processes become very large.  
Therefore, we would not find any solutions to satisfy the LFV constraint.
We will discuss only the case of NH in this section. 
Fig.~\ref{fig:tau} shows allowed region of $\tau$ in the fundamental space, where the blue color represents $\le1\sigma$, green 1-2$\sigma$, yellow 2-3$\sigma$, and red 3-5$\sigma$.

\begin{figure}[tb]\begin{center}
\includegraphics[width=80mm]{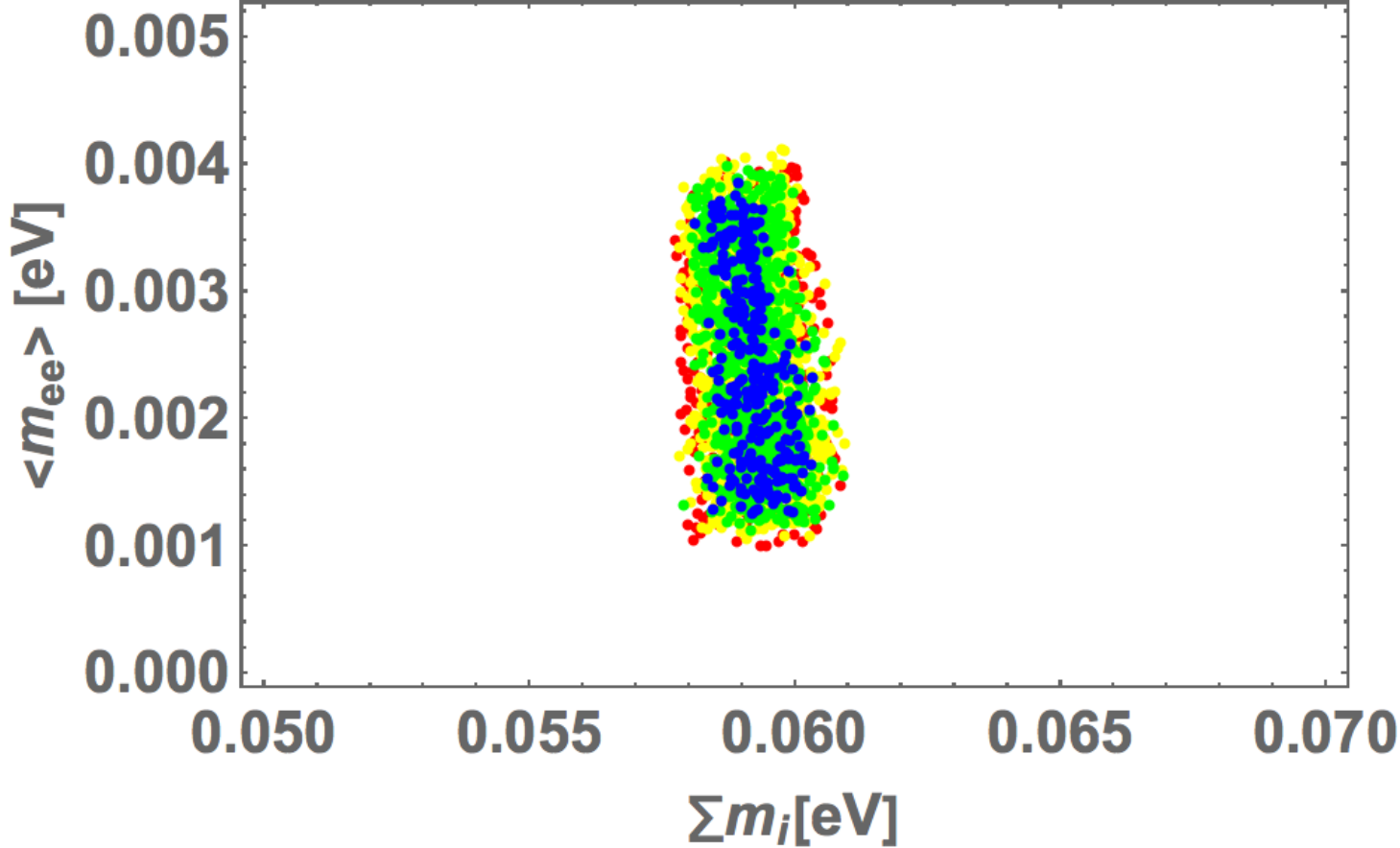}
\caption{The sum of neutrino masses $\sum m_i(\equiv$ Tr$[ D_\nu]$) versus the effective mass for the neutrinoless double beta decay $\langle m_{ee}\rangle$ in case of NH. The color distribution is the same as the one in Fig.~\ref{fig:tau}.}   
\label{fig:masses}\end{center}\end{figure}

Fig.~\ref{fig:masses} shows the sum of neutrino masses $\sum m_i(\equiv$ Tr$[ D_\nu]$) versus 
the effective mass for the neutrinoless double beta decay $\langle m_{ee}\rangle$, depending on the $\chi^2$ analysis. The color distribution is the same as the one in Fig.~\ref{fig:tau}.
It suggests that 1 meV $\lesssim \langle m_{ee}\rangle \lesssim$ 4 meV and  58 meV $\lesssim \sum m \lesssim$ 62 meV within the range of 5$\sigma$. It implies large neutrino mass hierarchies $m_{\nu_1}<<m_{\nu_2}<<m_{\nu_3}$, since the sum of masses is close to $\sqrt{\Delta m^2_{\rm atm}}$.
Notice here that the total neutrino masses are consistent with the recent cosmological constraint; $\sum m_i \le$ 120 meV.

\begin{figure}[tb]\begin{center}
\includegraphics[width=53mm]{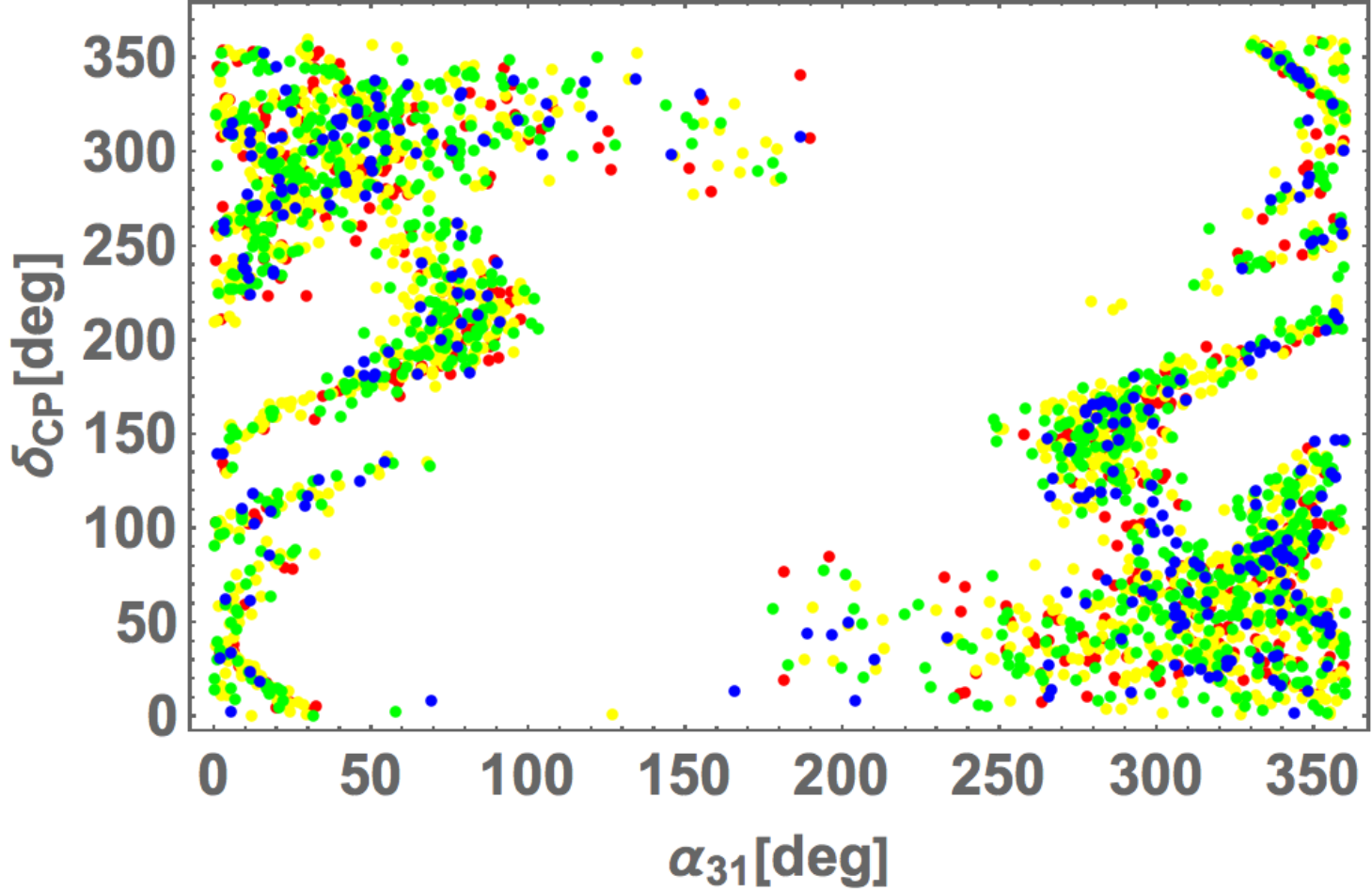}
\includegraphics[width=53mm]{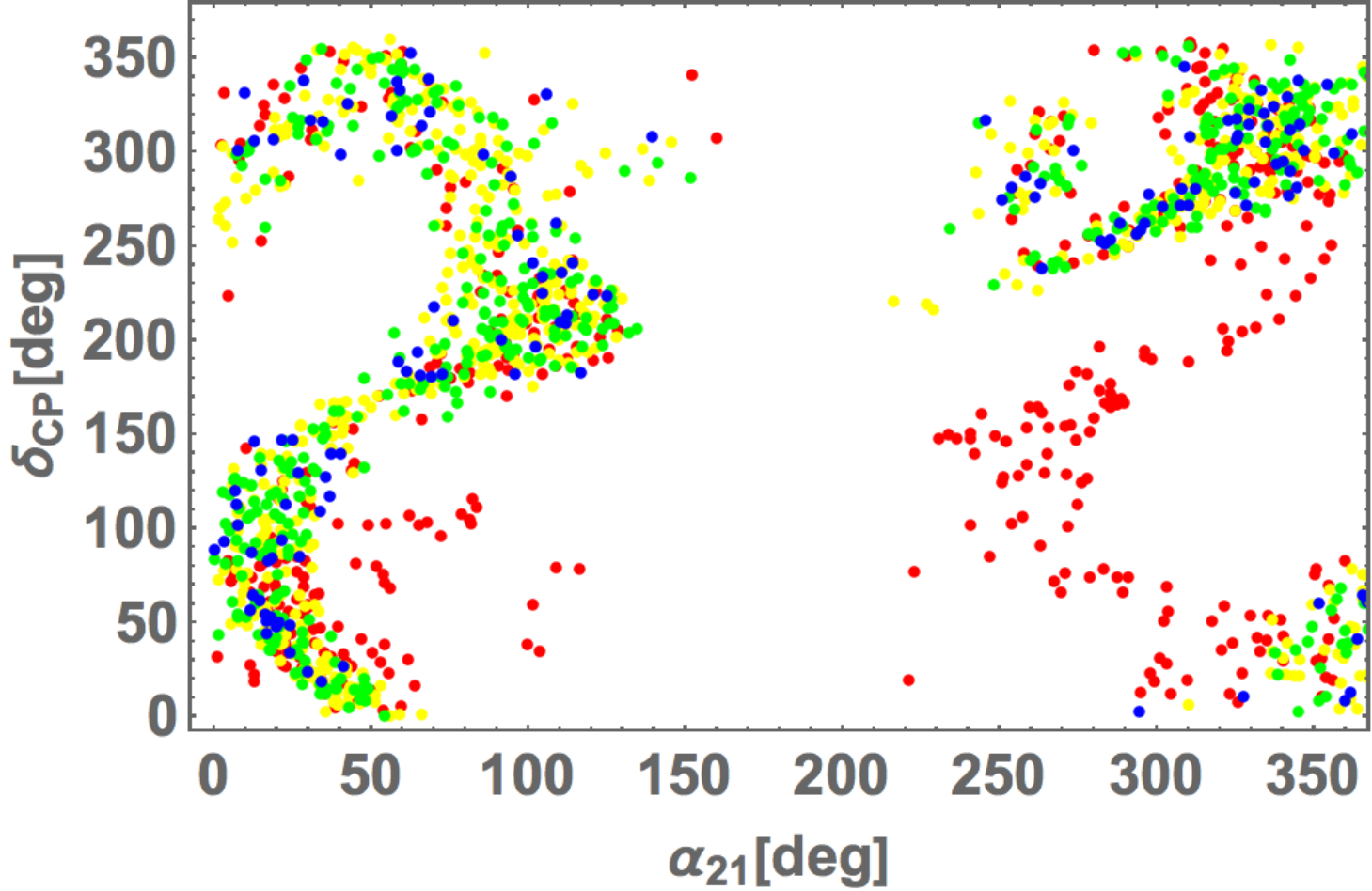}
\includegraphics[width=53mm]{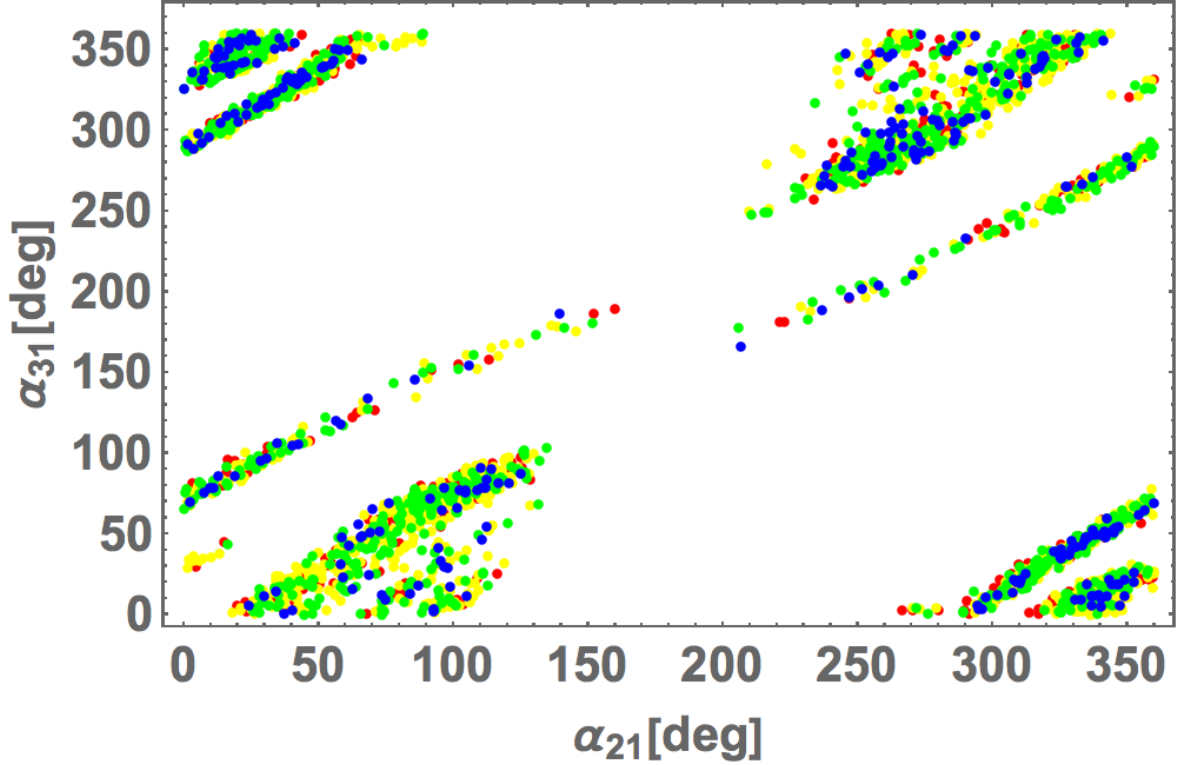}
\caption{
Correlations among Dirac CP phase $\delta_{CP}$ and Majorana phases $\alpha_{21,31}$ in case of NH. 
The color distribution is the same as the one in Fig.~\ref{fig:tau}.}
\label{fig:2}\end{center}\end{figure}

Fig.~\ref{fig:2} shows correlations among  Dirac CP phase $\delta_{CP}$ and Majorana phases $\alpha_{21,31}$ in case of NH. The color distribution is the same as the one in Fig.~\ref{fig:tau}.
The Dirac CP phase and the Majorana phase $\alpha_{31}$  run over whole the range. 
The Majorana phase  $\alpha_{21}$ runs over whole the range except around $180^\circ$.  
We show two sample points in Tab.~\ref{samplelepton},
where the first one is selected so as to minimize $\chi^2$, while the second one is chosen that  $\delta_{CP}^\ell$ be closest value of best fit; $195^\circ$.
In case of IH we have found  allowed region above the range of $2\sigma$, but all of the regions are excluded by the constraint of $\mu\to e\gamma$. Thus, we do not show the results in this case.

\begin{table}[tb]
	\centering
	\begin{tabular}{|c|c|c|} \hline 
		\rule[14pt]{0pt}{0pt}	  &  $\chi=0.975$    &    $\chi=1.54$ \\ \hline 
		\rule[14pt]{0pt}{0pt}	
		$\tau$&   $ -0.872142 + 1.58675 \, i$  & $-0.927 + 1.75\, i$    \\ 
		\rule[14pt]{0pt}{0pt}
		$[\frac{m_{0}}{\rm GeV},\frac{M_0}{\rm TeV},\frac{M_1}{\rm TeV}]$ 
		&$ [535, 373, 3393]$ & [ 536, 293, 2601]   \\
		\rule[14pt]{0pt}{0pt}
		$\lambda_{H\eta}''$ 
		&$ 1.44\times10^{-7} $ & $9.76\times10^{-8}$   \\
		\rule[14pt]{0pt}{0pt}
		$[\alpha_\ell,\beta_\ell,\gamma_\ell]$ 
		& $[0.00102, -0.354, 0.506]$ & $[0.00222, -0.320, 0.523]$   \\
		\rule[14pt]{0pt}{0pt} 
		$[\alpha_\eta,\gamma_\eta]$ 
		& $[0.00981e^{0.12i},  0.806e^{-3.13i}]$ &$[0.00958e^{-0.146i},  1.01e^{0.0105i}]$    \\
		\rule[14pt]{0pt}{0pt}
		$[\beta_{\eta_1},\beta_{\eta_2}]$ 
		& $[ 0.176e^{2.24i}, 0.455e^{-2.95i}]$ & $[ 0.128e^{-2.34i}, 0.638e^{-0.103i}]$    \\
		\rule[14pt]{0pt}{0pt}
		$V_{eL}$ 
		&
		${\tiny 
		\left(
   		 \begin{array}{ccc}
     		 0.325 & -0.108 & -0.940 \\
  	 	  	-0.0570 - 0.270 i& -0.1055 - 0.955 i& -0.00757+  0.0163 i \\
			0.173 - 0.888 i& -0.0765 + 0.243 i& 0.0684 - 0.335 i
   		 \end{array}
 		 \right)}$
		&
		${\tiny 
		\left(
   		 \begin{array}{ccc}
     		 0.243 & -0.0493 & -0.969 \\
-0.0192 - 0.164 i& -0.0506 - 0.985 i& -0.00223 +  0.00905 i \\
			0.108 - 0.950 i& -0.0287 + 0.155 i& 0.0284 - 0.246 i
   		 \end{array}
 		 \right)}$
		\\
		$V_{\nu}$ 
		&
		${\tiny 
		\left(
   		 \begin{array}{ccc}
     		 -0.492+ 0.178 i&0.467 + 0.465 i& -0.0362 + 0.539 i  \\
-0.175 - 0.131 i& -0.470 - 0.468 i& 0.130 + 0.704 i \\
			-0.630 + 0.530 i& -0.232 - 0.269 i& -0.286 - 0.338 i
			   		 \end{array}\right)}$
		&
		${\tiny 
		\left(
   		 \begin{array}{ccc}
     		-0.675 + 0.0560 i& 0.337 - 0.358 i& -0.467 + 0.286 i\\
  	 	  	-0.0264 - 0.114 i& 0.210 + 0.642 i& 0.122 + 0.718 i \\
			-0.313 + 0.655 i& 0.0919 + 0.541 i& -0.0671 - 0.408 i
   		 \end{array}
 		 \right)}$
		\\
		$\sin\theta_{12}$ & $ 0.548$	& $ 0.544$ \\
		\rule[14pt]{0pt}{0pt}
		$\sin\theta_{23}$ &  $ 0.756$	& $ 0.766$  \\
		\rule[14pt]{0pt}{0pt}
		$\sin\theta_{13}$ &  $ 0.148$	&  $ 0.150$ \\
		\rule[14pt]{0pt}{0pt}
		$\delta_{CP}^\ell$ &  $275^\circ$ 	&  $194^\circ$ 	 \\
		\rule[14pt]{0pt}{0pt}
		$[\alpha_{21},\,\alpha_{31}]$ &  $[251^\circ,\,336^\circ]$ 	&$[313^\circ,\,333^\circ]$ 	 \\	
		\rule[14pt]{0pt}{0pt}
		$\sum m_i$ &  $58.9$\,meV 	& $58.8$\,meV \\
		\rule[14pt]{0pt}{0pt}
		$\langle m_{ee} \rangle$ &  $2.51$\,meV 	&  $3.61$\,meV  \\
		\hline
	\end{tabular}
	\caption{Numerical values of parameters and observables
		at the sample points of NH.}
	\label{samplelepton}
\end{table}

\section{Conclusion and discussion}
\label{sec:conclusion}
We have explored a predictive lepton model with a modular $S_4$ symmetry, in which we have generated the neutrino mass matrix at one-loop level. The tree-level mass matrix is forbidden by well-assigned modular weights, which also play an important role in stabilizing dark matter candidate due to a remnant $Z_2$ symmetry even after breaking the modular symmetry. It implies one does not need to impose any additional symmetries such as $Z_2$ by hand.
Embedding three families of the Majorana neutrinos, right-handed charged-leptons and left-handed charged-leptons into singlet, doublet, and triplet respectively under $S_4$.
We have found allowed regions to satisfy all the relevant experimental constraints and obtained several predictions  for NH.
While we have not found the allowed region in case of IH, since whole the allowed region to satisfy the neutrino oscillation data conflicts with the constraint of $\mu\to e\gamma$. 
Then, we have shown our numerical results in figures such as phases, mixings, and neutrino masses, applying $\chi^2$ analysis.
We have also demonstrate two sample points, imposing on minimizing $\chi^2$ and best fit value of $\delta_{CP}^\ell$ of $195^\circ$.
We have found that 1 meV $\lesssim \langle m_{ee}\rangle \lesssim$ 4 meV and  58 meV $\lesssim \sum m \lesssim$ 62 meV within the range of 5$\sigma$. 
It suggests that large neutrino mass hierarchies $m_{\nu_1}<<m_{\nu_2}<<m_{\nu_3}$ is realzied, since the sum of masses is close to $\sqrt{\Delta m^2_{\rm atm}}$.
Notice here that the total neutrino masses are consistent with the recent cosmological constraint; $\sum m_i \le$ 120 meV.
\section*{Acknowledgments}
\vspace{0.5cm}
{\it
This research was supported by an appointment to the JRG Program at the APCTP through the Science and Technology Promotion Fund and Lottery Fund of the Korean Government. This was also supported by the Korean Local Governments - Gyeongsangbuk-do Province and Pohang City (H.O.). 
H. O. is sincerely grateful for the KIAS member, and log cabin at POSTECH to provide nice space to come up with this project.
Y. O. was supported from European Regional Development Fund-Project Engineering Applications of Microworld
Physics (No.CZ.02.1.01/0.0/0.0/16\_019/0000766)}

\appendix
\section{Mass structure}

$y_i$'s are given by
\begin{eqnarray}
y_1 &=& - 3 \pi \left(\frac{b_1}{8} + 3 b_5\right), \\
y_2 &=& 3 \sqrt{3} \pi b_3, \\ 
y_3 &=& - \pi \left( -\frac{b_1}{4} + 2 b_5\right), \\ 
y_4 &=& - \pi \sqrt{2} b_2, \\
y_5 &=& - 4\pi \sqrt{2} b_4, 
\end{eqnarray}
where $b_i$ are given by 
\begin{eqnarray}
b_1 \sim 1, \ 
b_2 \sim q, \ 
b_3 \sim q^2, \  
b_4 \sim 0, \ 
b_5 \sim 0, 
\end{eqnarray}
with $q=\exp( 2 \pi i \tau)$ and $|q| \ll 1$ \cite{Novichkov:2019sqv}.  


The structures of the majorana mass matrix and the yukawa coupling $y_\eta$ is written by 
\begin{eqnarray}
M_N = M_0
\left[
\begin{array}{ccc}
{\cal O}(q^0) & {\cal O}(q^2) & {\cal O}(q^0) \\
{\cal O}(q^2) & {\cal O}(q^0) & {\cal O}(q^2) \\
{\cal O}(q^0) & {\cal O}(q^2) & 0 \\
\end{array}
\right], 
\label{eq:majo}
\end{eqnarray}
\begin{eqnarray}
y_\eta =
\left[
\begin{array}{ccc}
{\cal O}(q^2) & {\cal O}(q^0) & {\cal O}(q^2) \\
{\cal O}(q^1) & {\cal O}(q^3) & {\cal O}(q^1) \\
{\cal O}(q^3) & {\cal O}(q^1) & {\cal O}(q^3) \\
\end{array}
\right],
\label{eq:yuketa}
\end{eqnarray}
where we assume $M_0 \sim M_1$. 
Using Eq.(\ref{eq:majo}), (\ref{eq:yuketa}) and (\ref{eq:numass}), we can obtain the neutrino mass structure as following: 
\begin{eqnarray}
m_\nu =
\left[
\begin{array}{ccc}
{\cal O}(q^0) & {\cal O}(q^3) & {\cal O}(q^1) \\
{\cal O}(q^3) & {\cal O}(q^2) & {\cal O}(q^4) \\
{\cal O}(q^1) & {\cal O}(q^4) & {\cal O}(q^2) \\
\end{array}
\right]. 
\end{eqnarray}
Two eigenvalues of $m_\nu$ are proportional to $q^2$ and one is proportional to $q^0$. 
It means that the model has two light neutrino masses and one heavy mass.   

\section{Confidence level}

We discuss the calculation of the CL. 

The probability density function is written by the following form: 
\begin{eqnarray}
 f(x, \nu)= \frac{x^{\nu/2-1} \exp (\frac{x}{2})}{2^{\nu/2} \Gamma (\frac{\nu}{2})},  
\end{eqnarray}
where $\nu$ is the degree of freedom (DOF). 
It is normalized by 
\begin{eqnarray}
 \int^\infty_0 f(x, \nu) dx = 1. 
\end{eqnarray}

The CL for $\nu$ DOF is given by 
\begin{eqnarray}
 \int_0^{\Delta \chi^2} f(x, \nu) dx,  
\end{eqnarray}
where $\Delta \chi^2 = \sum_{i=1}^{\nu} \left( \chi_i^2 - \chi^2_{i, min} \right)$ 
and all parameters are independent.
We can get the values $\chi_i^2$ and $\chi_{i, min}^2$ from NuFIT 5.0~\cite{Esteban:2020cvm}. 

\begin{table}[thb]
  \begin{tabular}{|c||c|c|c|c|c|c|}
  \hline
    CL(\%)                & $\nu =1$ & $\nu =2$ & $\nu =3$ & $\nu =4$ & $\nu =5$ & $\nu =6$ \\
  \hline
    68.27 (1$\sigma$)  & 1.00		  & 2.30		& 3.53		   & 4.72		& 5.89	    	  &	 7.04		\\
  \hline
    95.45 (2$\sigma$)  & 4.00		  & 6.18		& 8.02		   & 9.72		& 11.31     &	 12.85		\\
  \hline
    99.73 (3$\sigma$)  & 9.00		  & 11.83		& 14.16	   & 16.25		& 18.21   	  &	 20.06		\\
  \hline
    100$-$5.7$\times$10$^{-5}$ (5$\sigma$) 
     & 25.00		  & 28.74		& 31.81	   & 34.56		& 37.09   	  &	 39.49		\\
  \hline
\end{tabular}
\caption{Values $\Delta \chi^2$ corresponding to CL for joint estimation of $\nu$ parameters.}
\label{tab:chiCL}
\end{table}
Tab.~\ref{tab:chiCL} shows values $\Delta \chi^2$ corresponding to CL for joint estimation of $\nu$ parameters.

\end{document}